\documentclass{article}

\def\xxinput#1{\input#1}

\xxinput{vsolj02.sty}

\usepackage{longtable}
\usepackage[dvipdfmx]{graphicx}
\usepackage[comma,colon]{natbib}

\usepackage[OT2,T1]{fontenc}

\def\cite{\citealt}
\setcitestyle{aysep={}}

\def\commenta{$^*$}
\def\commentb{$^\dagger$}
\def\commentc{$^\ddagger$}
\def\commentd{$^\S$}
\def\commente{$^\|$}
\def\commentf{$^\#$}

\newcounter{author}
\def\altaffilmark#1{$^{#1}$}
\def\altaffiltext#1{$^{#1}$\,}

\def\authorcount#1#2{{\refstepcounter{author}\label{#1}
                     \altaffiltext{\ref{#1}}{#2}}}

\def\Shibataprep{M. Shibata et al. in preparation}

\begin{document}

\begin{center}

\title{Analysis of the IW And star ASAS J071404$+$7004.3}

\author{
        Taichi~Kato,\altaffilmark{\ref{affil:Kyoto}}
        Kiyoshi~Kasai,\altaffilmark{\ref{affil:Kai}}
        Elena~P.~Pavlenko,\altaffilmark{\ref{affil:CrAO}}$^,$\altaffilmark{\ref{affil:CrimeanFU}}
        Nikolaj~V.~Pit,\altaffilmark{\ref{affil:CrAO}}
        Aleksei~A.~Sosnovskij,\altaffilmark{\ref{affil:CrAO}}
}
\vskip -2mm
\author{
        Hiroshi~Itoh,\altaffilmark{\ref{affil:Ioh}}
        Hidehiko~Akazawa,\altaffilmark{\ref{affil:Aka}}
        Stephen~M.~Brincat,\altaffilmark{\ref{affil:Brincat}}
        Leonid~E.~Keir,\altaffilmark{\ref{affil:Mayaki}}
        Sergei~N.~Udovichenko,\altaffilmark{\ref{affil:Mayaki}}
}
\vskip -2mm
\author{
        Yusuke~Tampo,\altaffilmark{\ref{affil:Kyoto}}
        Naoto~Kojiguchi,\altaffilmark{\ref{affil:Kyoto}}
        Masaaki~Shibata,\altaffilmark{\ref{affil:Kyoto}}
        Yasuyuki~Wakamatsu,\altaffilmark{\ref{affil:Kyoto}}
        Tam\'as~Tordai,\altaffilmark{\ref{affil:Polaris}}
}
\vskip -2mm
\author{
        Tonny~Vanmunster,\altaffilmark{\ref{affil:Vanmunster}}
        Charles~Galdies\altaffilmark{\ref{affil:Caldies}}
}
\email{tkato@kusastro.kyoto-u.ac.jp}

\authorcount{affil:Kyoto}{
     Department of Astronomy, Kyoto University, Sakyo-ku,
     Kyoto 606-8502, Japan}

\authorcount{affil:Kai}{
     Baselstrasse 133D, CH-4132 Muttenz, Switzerland}

\authorcount{affil:CrAO}{
     Federal State Budget Scientific Institution ``Crimean Astrophysical
     Observatory of RAS'', Nauchny, 298409, Republic of Crimea}

\authorcount{affil:CrimeanFU}{
     V. I. Vernadsky Crimean Federal University, 4 Vernadskogo Prospekt,
     Simferopol, 295007, Republic of Crimea}

\authorcount{affil:Ioh}{
     Variable Star Observers League in Japan (VSOLJ),
     1001-105 Nishiterakata, Hachioji, Tokyo 192-0153, Japan}

\authorcount{affil:Aka}{
     Akazawa Funao Observartory, 107 Funao, Funaocho, Kurashiki,
     Okayama 710-0261, Japan}

\authorcount{affil:Brincat}{
     Flarestar Observatory, San Gwann SGN 3160, Malta}

\authorcount{affil:Mayaki}{
     Astronomical Observatory of Odessa I. I. Mechnikov National University,
     Marazlievskaya 1v, 65014 Odessa, Ukraine}

\authorcount{affil:Polaris}{
     Polaris Observatory, Hungarian Astronomical Association,
     Laborc utca 2/c, 1037 Budapest, Hungary}

\authorcount{affil:Vanmunster}{
     Center for Backyard Astrophysics Belgium, Walhostraat 1A,
     B-3401 Landen, Belgium}

\authorcount{affil:Caldies}{
     Institute of Earth Systems, University of Malta, Msida, MSD 2080, Malta}

\end{center}

\begin{abstract}
\xxinput{abst.inc}
\end{abstract}

\section{Introduction}

\citet{ini22j0714} reported on the recently identified
nearby, bright ($V \sim$ 12) cataclysmic variable (CV)
ASAS J071404$+$7004.3.
This object was identified as the optical counterpart
of the X-ray source 1RXS J071404.0$+$700413
by \citet{kir13asasrosatvar}
by examining light variations of ROSAT All Sky Survey
sources \citep{ROSATRXP} using the northern twin (ASAS-N)
of the All-Sky Automated Survey (ASAS: \cite{ASAS3}).
This object was unique among the objects identified
by \citet{kir13asasrosatvar} in that it was classified
as a miscellaneous variable star with a period of 22.44~d.

\citet{ini22j0714} published time-resolved optical
spectroscopy, X-ray observations and long- and short-term
optical variations.  The object was confirmed to be a CV
with an orbital period of 0.1366454(1)~d.
\citet{ini22j0714} compared the long-term light curve of
this object with the AAVSO light curves of prototypical
objects of CV subtypes: Z Cam (Z Cam star),
V513 Cas (IW And star) and VY Scl (VY Scl star) and
concluded that the object belongs
to the VY Scl-type CVs based on the visual similarity
of the light curve rather than detailed comparisons.

This object was introduced to us by a baavss-alert
message ``Request for help in monitoring ASAS J071404$+$\hspace{0pt}7004.3''
requested by Boris G{\"a}nsicke and posted by Jeremy Shears
on 2020 February 7\footnote{
   Although baavss-alert mailing list on yahoo had a publicly
   accessible archive, this service has already been
   discontinued and the message is not currently publicly
   available.
}.
Upon this message, one of the authors (TK) examined
All-Sky Automated Survey for Supernovae (ASAS-SN)
Sky Patrol data \citep{ASASSN,koc17ASASSNLC}
and readily identified it as an IW And star
(T. Kato, vsnet-alert 23944\footnote{
  $<$http://ooruri.kusastro.kyoto-u.ac.jp/mailarchive/vsnet-alert/23944$>$.
} on 2020 February 7) based on the heartbeat-type variation
with shallow dips in 2017 and the presence of standstills.
This identification was quickly reflected on
the AAVSO Variable Star Index \citep[VSX:][]{wat06VSX}\footnote{
  $<$https://www.aavso.org/vsx/index.php?view=detail.top\&oid=300502$>$.
}.  It is strange that there was no trace
of examination of this type identification in
\citet{ini22j0714} and we clarify the reasons of
the apparent discrepancy in the type classification.

In the following section, we describe the results from
ASAS-SN observations.  We conducted our own campaign of
time-resolved photometry by the VSNET Collaboration
\citep{VSNET}, which was aimed to detect possible negative
superhumps (T. Kato, vsnet-alert 23949\footnote{
  $<$http://ooruri.kusastro.kyoto-u.ac.jp/mailarchive/vsnet-alert/23949$>$.
}) motivated by the presence of negative superhumps
(a tilted disk) in the IW And star KIC 9406652
\citep{gie13j1922,kim20kic9406652,kim21kic9406652} and
a possible suggestion that a tilted disk might be responsible for
the IW And-type phenomenon \citep{kim20kic9406652}.
The log of observations is shown in table \ref{tab:log}.
The $R$ and $I$ bands refer to Kron-Cousins systems.
We also analyzed Transiting Exoplanet Survey Satellite (TESS)
observations.\footnote{
  $<$https://tess.mit.edu/observations/$>$.
}  The full light-curve
is available at the Mikulski Archive for Space Telescope
(MAST\footnote{
  $<$http://archive.stsci.edu/$>$.
}).

\begin{center}
\begin{longtable}{ccccccc}
\caption{Log of observations of ASAS J071404$+$7004.3.}\label{tab:log} \\
\hline
Start\commenta & End\commenta & mag\commentb & error\commentc &
$N$\commentd & obs\commente & band\commentf \\
\hline
\endfirsthead
\caption{Log of observations of ASAS J071404$+$7004.3 (continued).} \\
\hline
Start\commenta & End\commenta & mag\commentb & error\commentc &
$N$\commentd & obs\commente & band\commentf \\
\hline
\endhead
\hline
\multicolumn{7}{l}{\commenta JD$-$2400000.} \\
\multicolumn{7}{l}{\commentb Mean magnitude.} \\
\multicolumn{7}{l}{\commentc 1$\sigma$ of the mean magnitude.} \\
\multicolumn{7}{l}{\commentd Number of observations.} \\
\multicolumn{7}{l}{\commente Observer's code: Aka (H. Akazawa), BSM (S. Brincat),} \\
\multicolumn{7}{l}{CRI (Crimean Astrophys. Obs.), GCH (C. Galdies), Ioh (H. Itoh), } \\
\multicolumn{7}{l}{KU1 (Kyoto U. rooftop obs.), Kai (K. Kasai), May (Mayaki Obs.),} \\
\multicolumn{7}{l}{Trt (T. Tordai), Van (T. Vanmunster).} \\
\multicolumn{7}{l}{\commentf Filter. ``C'' means unfiltered.} \\
\\
\endfoot
\xxinput{obslog.inc}
\hline
\end{longtable}
\end{center}

\section{Results and discussions}

\subsection{Long-term light curve}

\begin{figure*}
\begin{center}
\includegraphics[width=16cm]{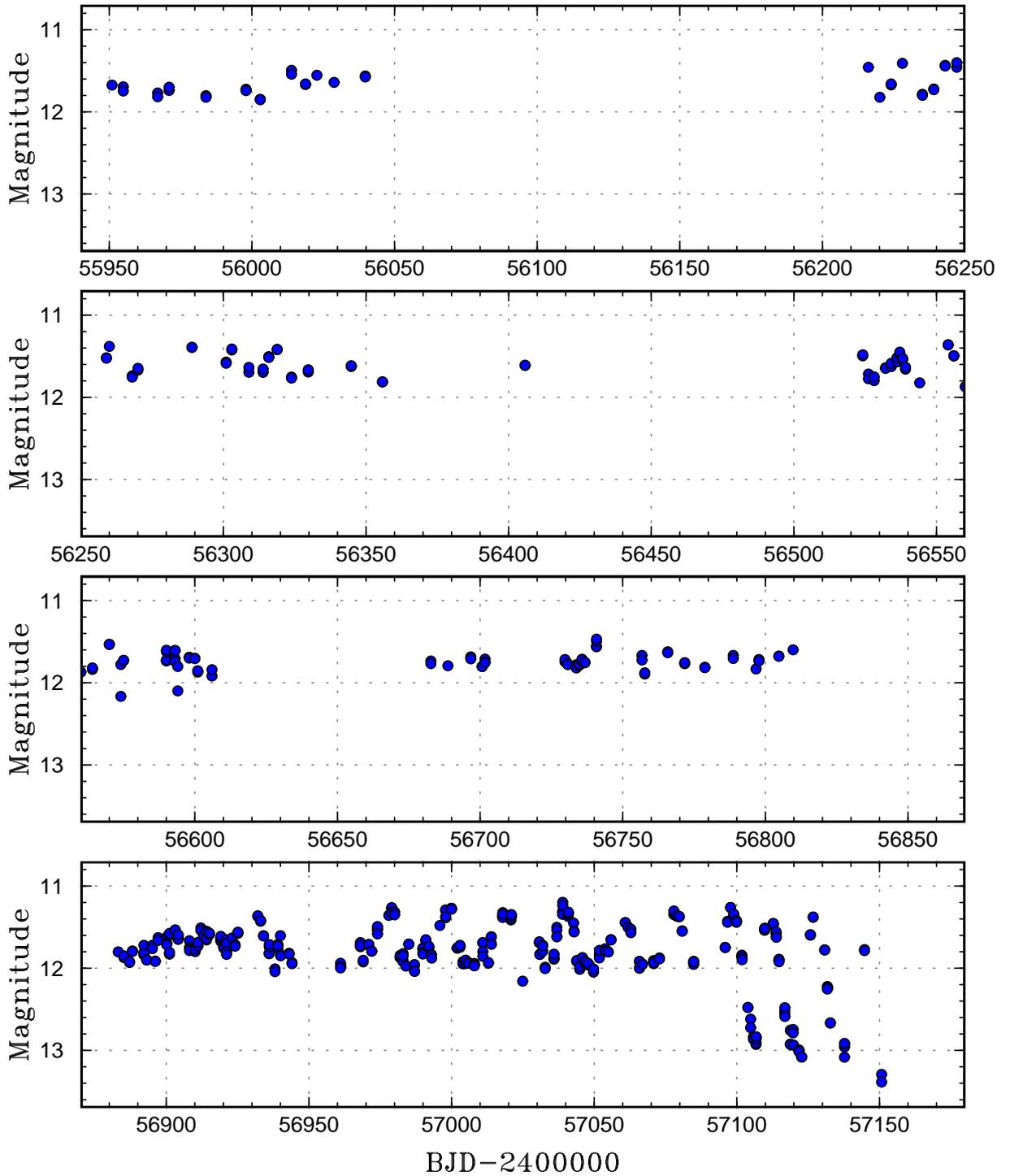}
\caption{
Long-term light curve of ASAS J071404$+$7004.3 using
the ASAS-SN data.  All data are in $V$ band.
After BJD 2456950, low-amplitude (0.7~mag) dwarf-nova
oscillations appeared.  After BJD 2457100, the amplitudes
of these oscillations became larger (1.6~mag) and
the period became shorter.
}
\label{fig:j0714lc}
\end{center}
\end{figure*}

\begin{figure*}
\begin{center}
\includegraphics[width=16cm]{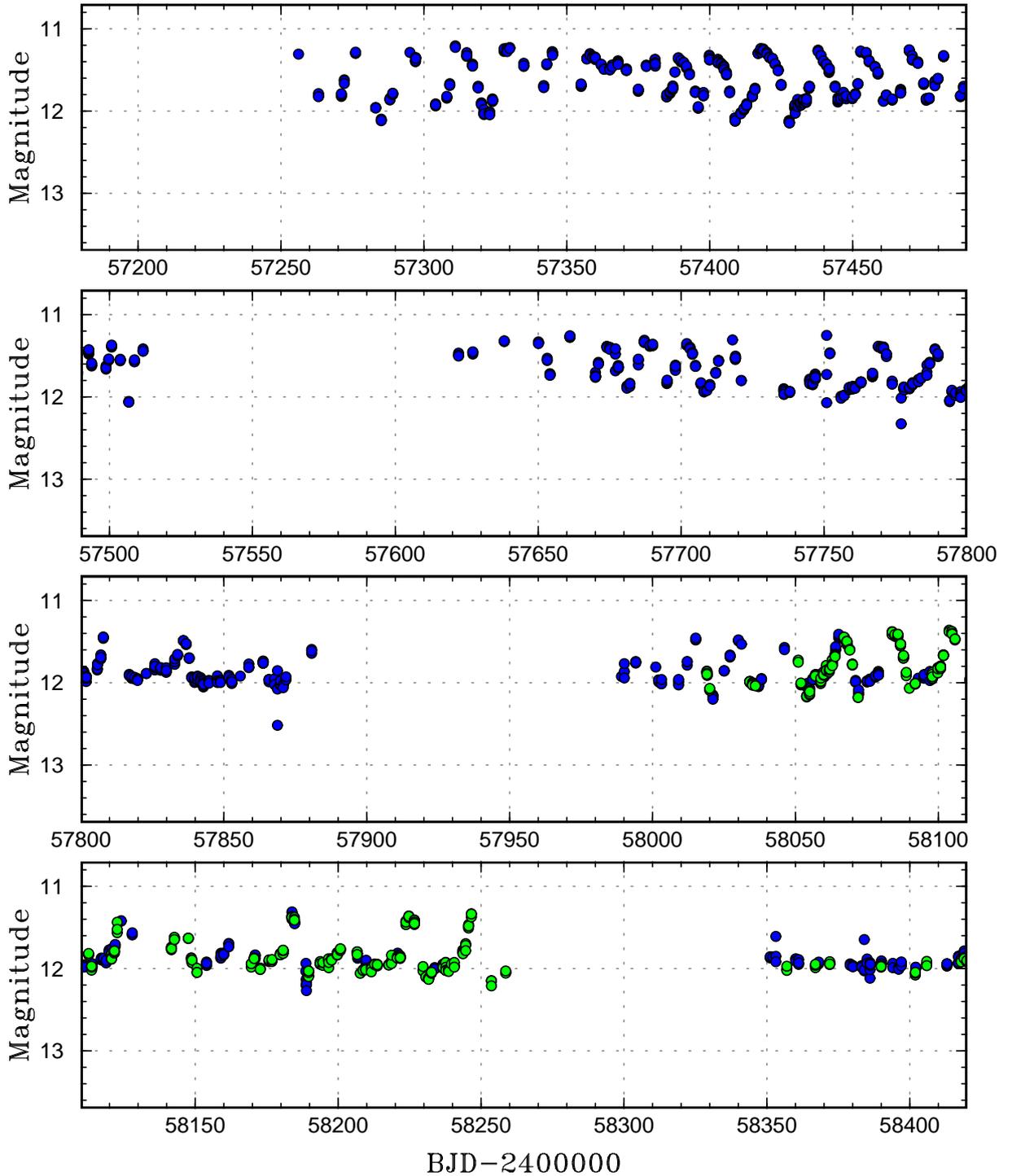}
\caption{
Long-term light curve of ASAS J071404$+$7004.3 using
the ASAS-SN data (2).  Blue and green symbols represent
$V$ and $g$ observations, respectively.
This segment is dominated by low-amplitude dwarf nova-type
variations with variable amplitudes and periods.
The periods became as short as 10~d.  In the latter
part of this segment, the periods became longer
and the rise to outbursts became slower than the decline.
Some outbursts were followed by small dips.
This part corresponds to the IW And-type behavior.
See figure \ref{fig:hopuplc2} for a comparison
to the IW And star HO Pup.
}
\label{fig:j0714lc2}
\end{center}
\end{figure*}

\begin{figure*}
\begin{center}
\includegraphics[width=16cm]{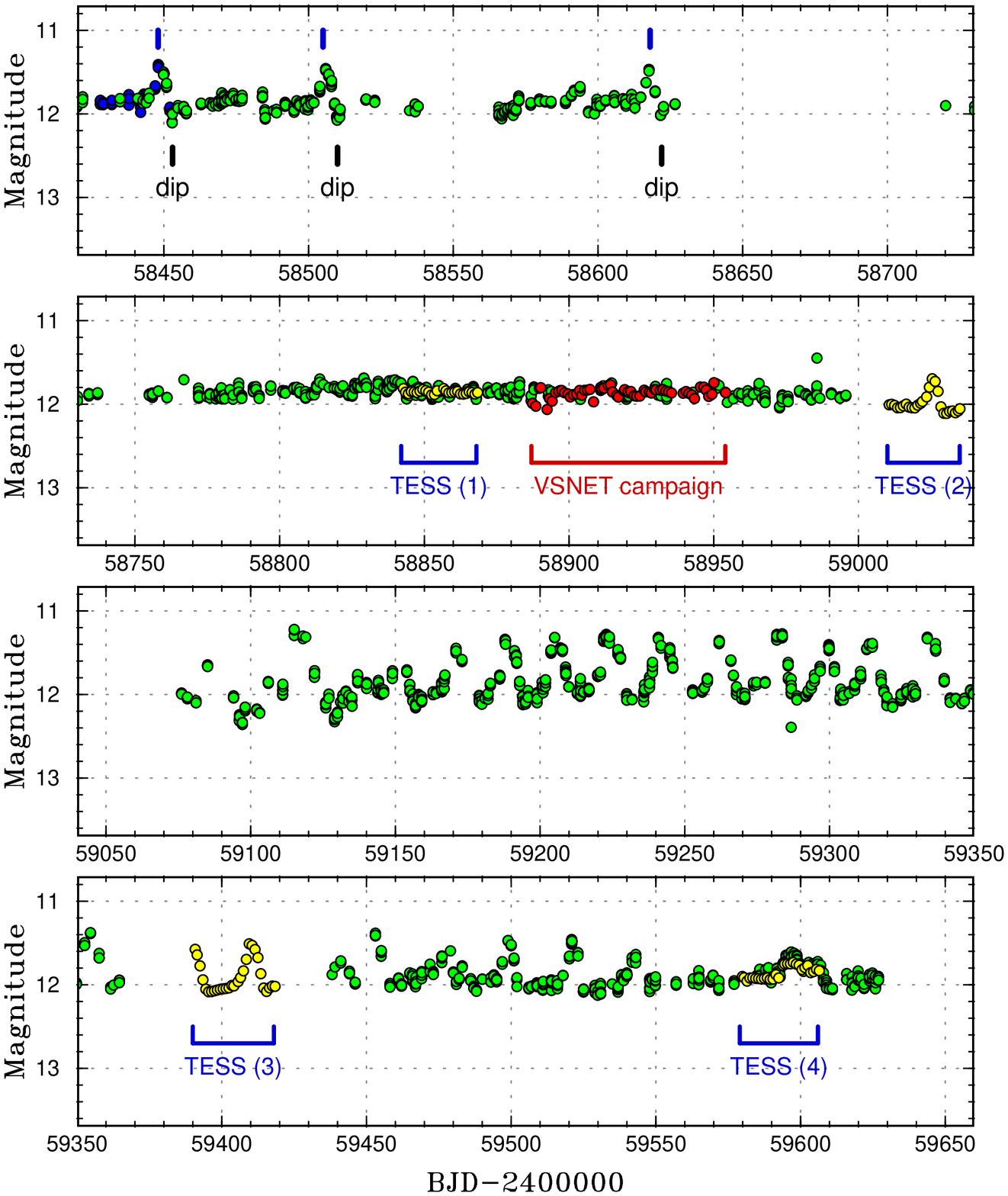}
\caption{
Long-term light curve of ASAS J071404$+$7004.3 using
the ASAS-SN data, TESS data (yellow) and the VSNET campaign
data (red).  Blue and green symbols
for the ASAS-SN data represent $V$ and $g$ observations,
respectively.
The TESS and VSNET data were binned to 1~d.
The initial three runs of of the TESS data were shifted
by $-$0.1 mag to match the ASAS-SN data.
The first panel shows the typical behavior of
an IW And star.  Slowly rising standstills were terminated
by outbursts (shown by the upper blue ticks),
followed by small dips (shown by the lower ticks).
See figure \ref{fig:hopuplc2} for a comparison
to the IW And star HO Pup.
The second panel shows the standstill phase.
An outburst in the TESS data (2) marked the end
of the standstill phase.
The third panel is also dominated by dwarf nova-type
variations.
}
\label{fig:j0714lc3}
\end{center}
\end{figure*}

Although the ASAS-SN light curves were shown in
\citet{ini22j0714}, they were either too small
(their figure 12) or the important feature was
not shown (their figure 2), we provide the full
ASAS-SN light curve
(figures \ref{fig:j0714lc}, \ref{fig:j0714lc2}
and \ref{fig:j0714lc3}) together with TESS observations
and the VSNET campaign data to show important features.

The most obvious feature is the long-lasting standstill
between BJD 2458720 and 2459000 (figure \ref{fig:j0714lc3},
the second panel).  It was rather a pity that the start
of this standstill was not recorded.  This standstill
was terminated by an outburst which occurred during
TESS observations.

The other parts after BJD 2456950 were dominated by
low-amplitude dwarf nova-type variations
(starting from the fourth panel of figure \ref{fig:j0714lc},
most of figure \ref{fig:j0714lc2} except the final
part and figure \ref{fig:j0714lc3} except the second
panel and after BJD 2459550).  The 22.44~d period
in \citet{kir13asasrosatvar} apparently referred to these
low-amplitude dwarf nova-type variations.
Although such low-amplitude dwarf nova-type variations are
also seen in IX Vel \citep{kat21ixvel}, the cycle length
in ASAS J071404$+$7004.3 is longer compared to
IX Vel with 13--20~d and the profile is different
in that variations in ASAS J071404$+$7004.3 tend to have
slower rise and steeper decline as seen in
the third panel of figure \ref{fig:j0714lc2}.
This feature was interpreted as ``inside out'' (outburst)
in \citet{ini22j0714}.

\subsection{Problem of type classification in \citet{ini22j0714}}

\begin{figure*}
\begin{center}
\includegraphics[width=16cm]{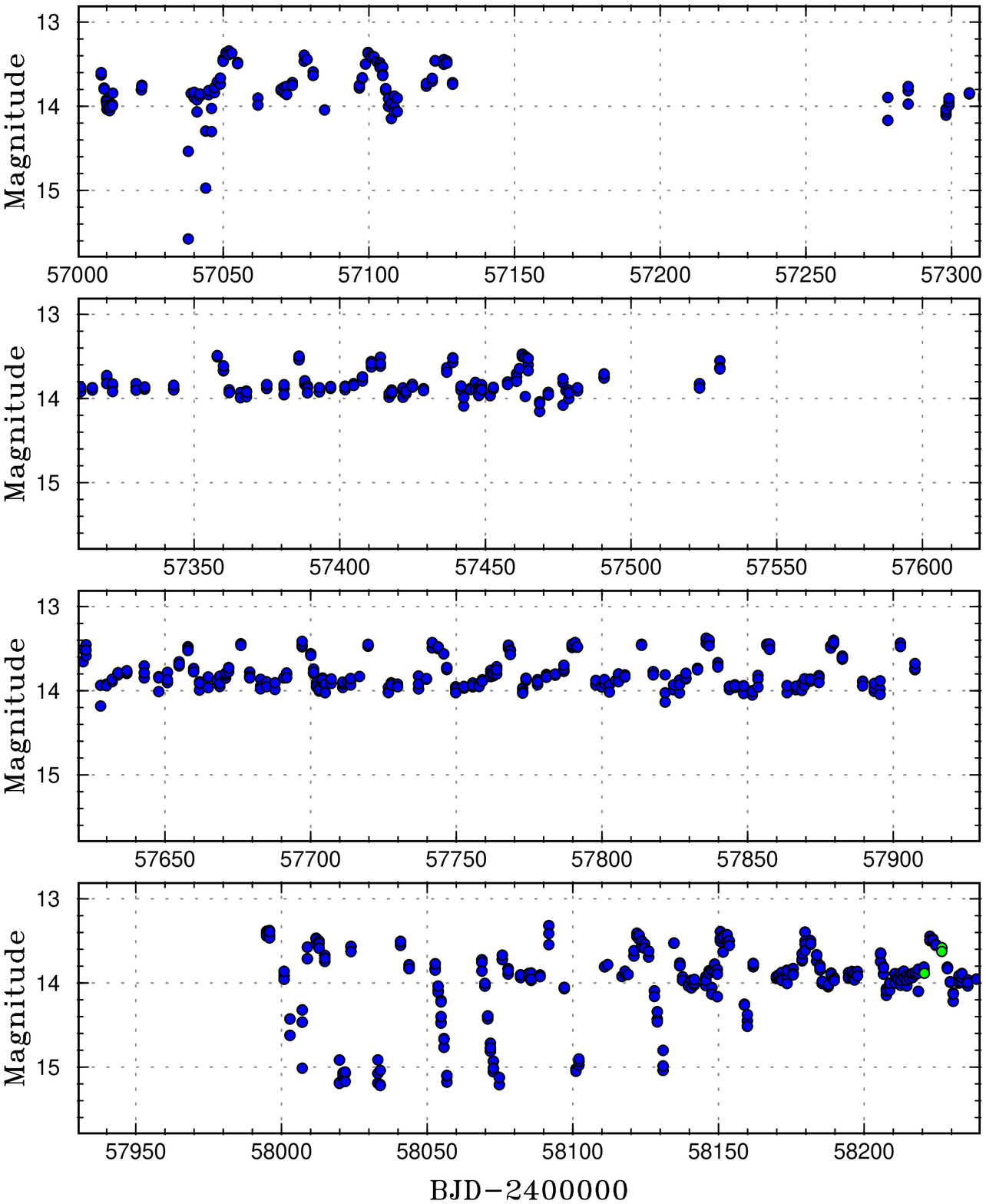}
\caption{
Long-term light curve of HO Pup using
the ASAS-SN data.  Blue and green symbols represent
$V$ and $g$ observations, respectively.
The third panel corresponds to ``heartbeat-type''
oscillations in IW And stars, accompanied by
slower rise and steeper decline.  This type of
variation is considered to be IW And-type cycles
simply lacking post-outburst dips.
After BJD 2458080, typical IW And-type behavior
is seen: standstills terminated by outbursts, followed by
dips.  It is also apparent that the depths of the dips
vary and the dips became inapparent after BJD 2458180,
indicating that heartbeat-type oscillations and
the typical IW And-type behavior form a smooth continuum.
}
\label{fig:hopuplc}
\end{center}
\end{figure*}

\begin{figure*}
\begin{center}
\includegraphics[width=16cm]{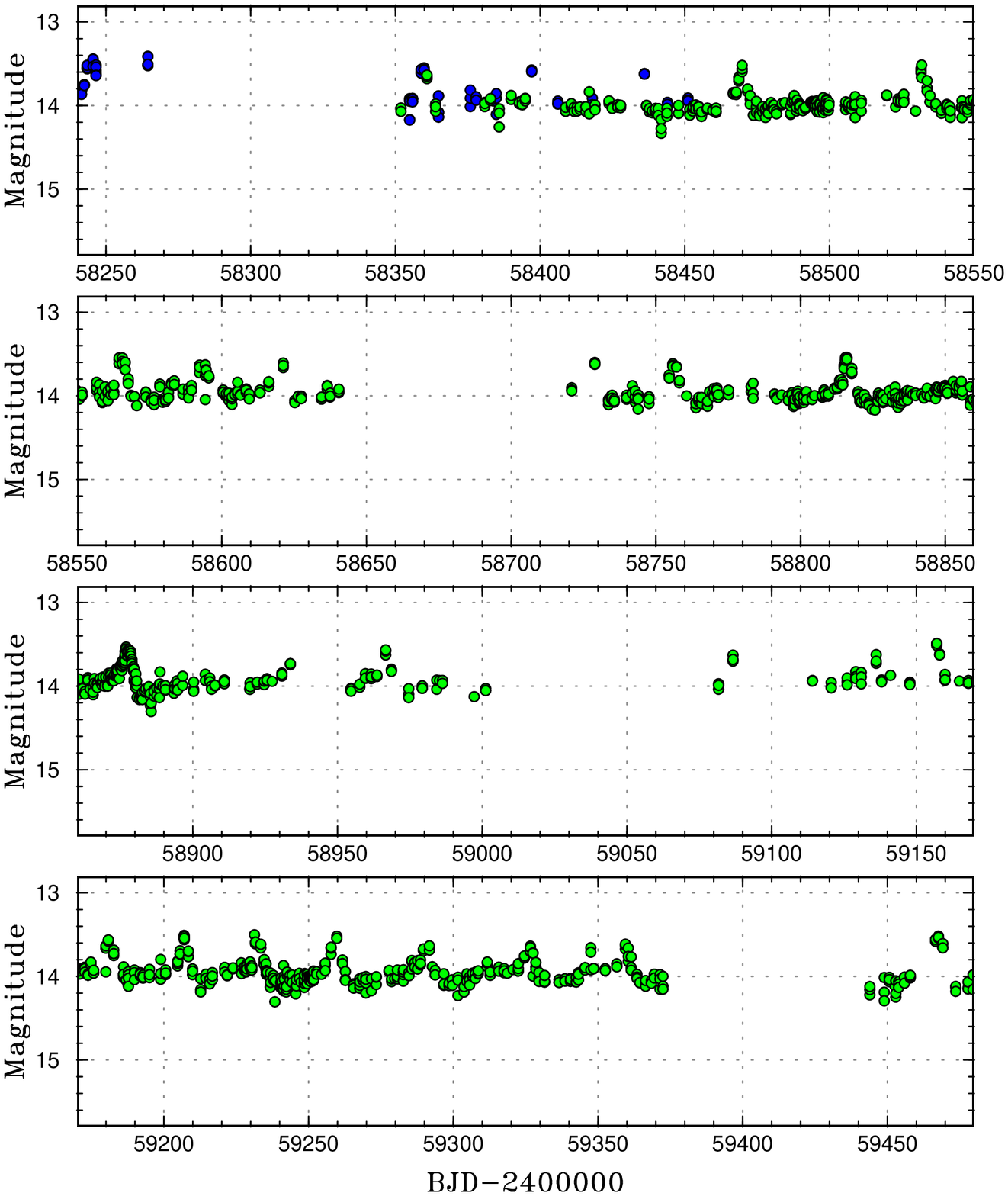}
\caption{
Long-term light curve of HO Pup using
the ASAS-SN data (2).  Blue and green symbols represent
$V$ and $g$ observations, respectively.
Most of the time, this star showed heartbeat-type oscillations
with variable intervals.  This segment is very similar to
ASAS J071404$+$7004.3 between BJD 2458150 and 2458630
(figures \ref{fig:j0714lc2} and \ref{fig:j0714lc3}).
}
\label{fig:hopuplc2}
\end{center}
\end{figure*}

The main reason why \citet{ini22j0714} reached
an inadequate classification was that they
only consulted the relatively old literature \citet{ham14zcam}
when referring to IW And stars.
It is queer that \citet{ini22j0714} referred to the category
IW And stars, which term was not mentioned
in \citet{ham14zcam}, but was later defined
by \citet{kat19iwandtype}.  These authors apparently
dismissed this later definition and subsequent progress
\citep[e.g.][]{kim20iwandmodel,kat20imeri,kat21bocet}.
In \citet{kat19iwandtype,kim20iwandmodel}, the IW And stars
were defined to have the features: (1) Standstills are
terminated by brightening in IW And-type objects,
unlike fading in Z Cam stars. (2) There are quasi-periodic cycles
consisting of a (quasi-)standstill with damping oscillations --- 
brightening which terminates the standstill --- often a deep dip
(not always present within the same object) and returning
to a (quasi-)standstill.
When deep dips are not present, the light curve looks
like a periodic pattern with gradual rise with
an increasing rate to a maximum followed by rapid fading.
This pattern is referred
to as ``heartbeat-type oscillations''.
It can be confirmed that this is a variety of IW And-type
variation since the same object shows both typical IW And-type
cycles with deep dips and these heartbeat-type oscillations.
The depths of the dips form a continuum from zero
(heartbeat-type oscillations) to a few magnitudes
in the typical IW And-type cycles.
This feature was described for the two objects
in FY Vul and HO Pup in \citet{kim20iwandmodel}, and
\citet{ini22j0714} probably missed modern knowledge of
IW And stars.  Considering these features, the light curve of
ASAS J071404$+$7004.3 is very similar to that of HO Pup
(figures \ref{fig:hopuplc} and \ref{fig:hopuplc2}).
Just for information, \citet{lee21hopup}
correctly described HO Pup as an IW And star.
The presence of long standstills (second panel in
figure \ref{fig:j0714lc3} and possibly before
BJD 2456930 in figure \ref{fig:j0714lc})
is also similar to the IW And star BO Cet,
which showed a long standstill (this object was therefore
initially classified as a novalike)
lasting for years \citep{kat21bocet}.

The second reason is that \citet{ini22j0714} did not
describe the characteristics of subclasses of CVs correctly.
\citet{ini22j0714} described IW And stars as showing
continual outbursts (not unlike dwarf novae) during
the high state (i.e. during the ``standstill'').
The most obvious feature of the IW And stars, however, is
the presence of outbursts at the end of standstills,
not during standstills.

Low states in VY Scl stars were also poorly defined
(without a reference) and poorly compared with
ASAS J071404$+$7004.3 in \citet{ini22j0714}.
Typical low states of VY Scl stars lasts months to
years and there are usually no strong sign of
dwarf-nova outbursts during decline despite that
the mass-transfer rates decrease.
Examples of the light curves and a schematic
representation of the lack of dwarf-nova outbursts
can be found in \citet{lea99vyscl}.  This feature
has long been known and the cause of the apparent
inconsistency with the disk-instability model
has been sought.  \citet{lea99vyscl} suggested
heating from a hot white dwarf suppresses
thermal instability of the disk.
\citet{ham02vysclmagnetic} considered that
the magnetism of a white dwarf could truncate
the disk, thereby suppressing thermal instability.
The presence of magnetism of white dwarfs in
VY Scl stars in general, however, has not been
apparent by observations.
In any case, \citet{ini22j0714} did not consider
this well-known observational feature of VY Scl stars.
The fading episode in ASAS J071404$+$7004.3
was simply a dwarf nova phase with larger amplitudes
(as in Z Cam stars) when the mass-transfer rate
dropped slightly.  It is also apparent from figure
\ref{fig:hopuplc} that excursions to fainter magnitudes
while maintaining dwarf nova-type outbursts happen
in other IW And stars.  This is probably what happened
ASAS J071404$+$7004.3 after BJD 2457100.

The confusion with the VY Scl-type phenomenon is not,
however, surprising.  Even the authority of CVs Brian Warner
considered dwarf nova-type AM CVn stars, which are
currently known as helium counterpart of SU UMa/WZ Sge-type 
stars in hydrogen-rich CVs
\citep[see e.g.][]{tsu97amcvn,kot12amcvnoutburst},
to be analogous to VY Scl stars at least in the past
\citep{war95amcvn}.

\subsection{Analysis of ground-based time-resolved campaign}

\begin{figure*}
\begin{center}
\includegraphics[width=16cm]{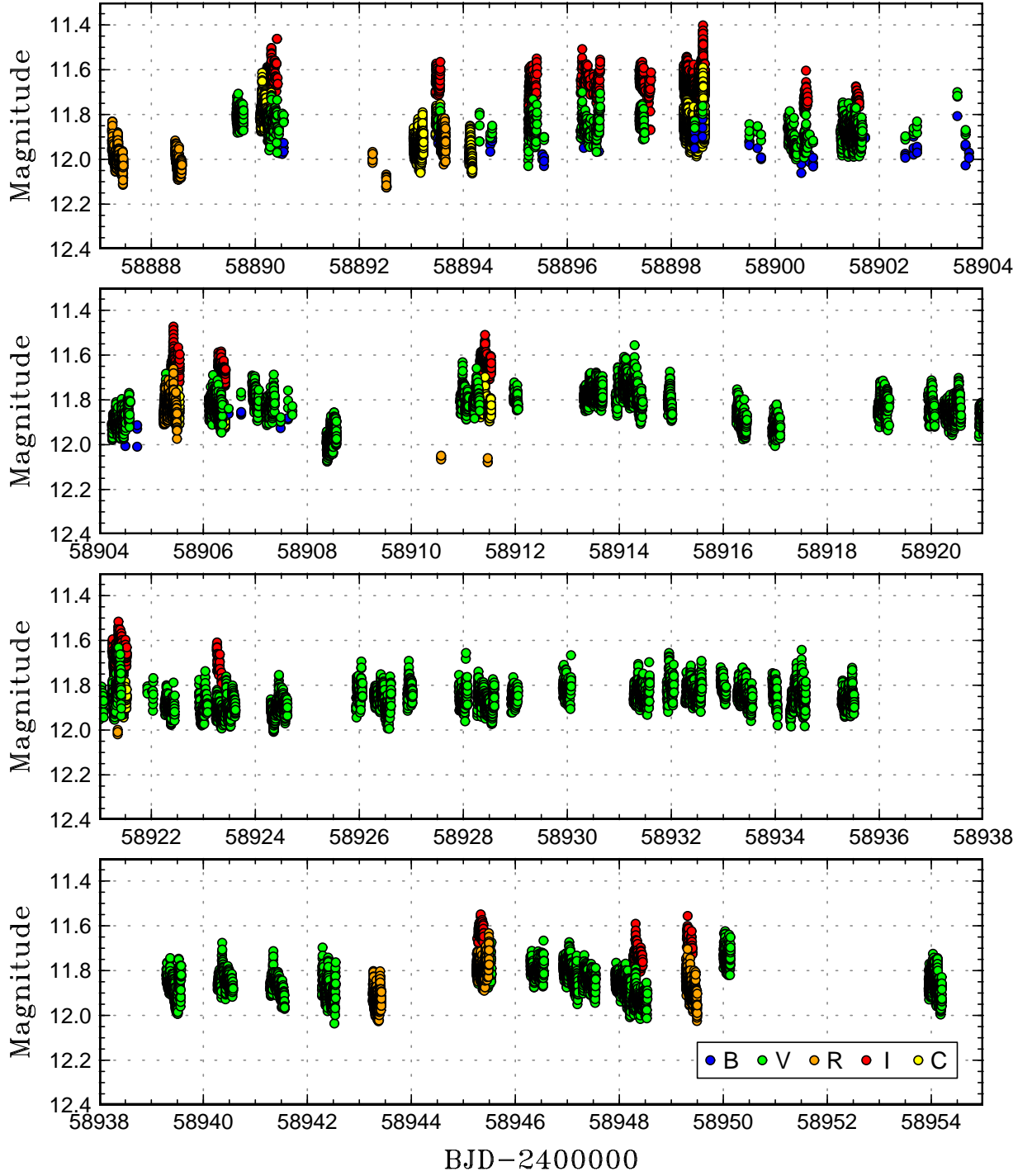}
\caption{
Light curve of ASAS J071404$+$7004.3 using the ground-based
observations obtained by our campaign.
The zero points were adjusted by correcting the differences
between observers and bands.  The color indices against $V$
were determined only for $B$ and $I$ by simultaneous
observations, and they are reflected on this figure.
The colors corresponded to $B-V$=$+$0.08 and $V-I$=$+$0.19.
We assumed that the mean magnitudes were the same
for $V$, $R$ and C since there were no simultaneous data
for these bands and since the color indices between
these bands are expected to be close to zero in
this unreddened and disk-dominated CV.
}
\label{fig:j0714vsnet}
\end{center}
\end{figure*}

\begin{figure*}
\begin{center}
\includegraphics[width=16cm]{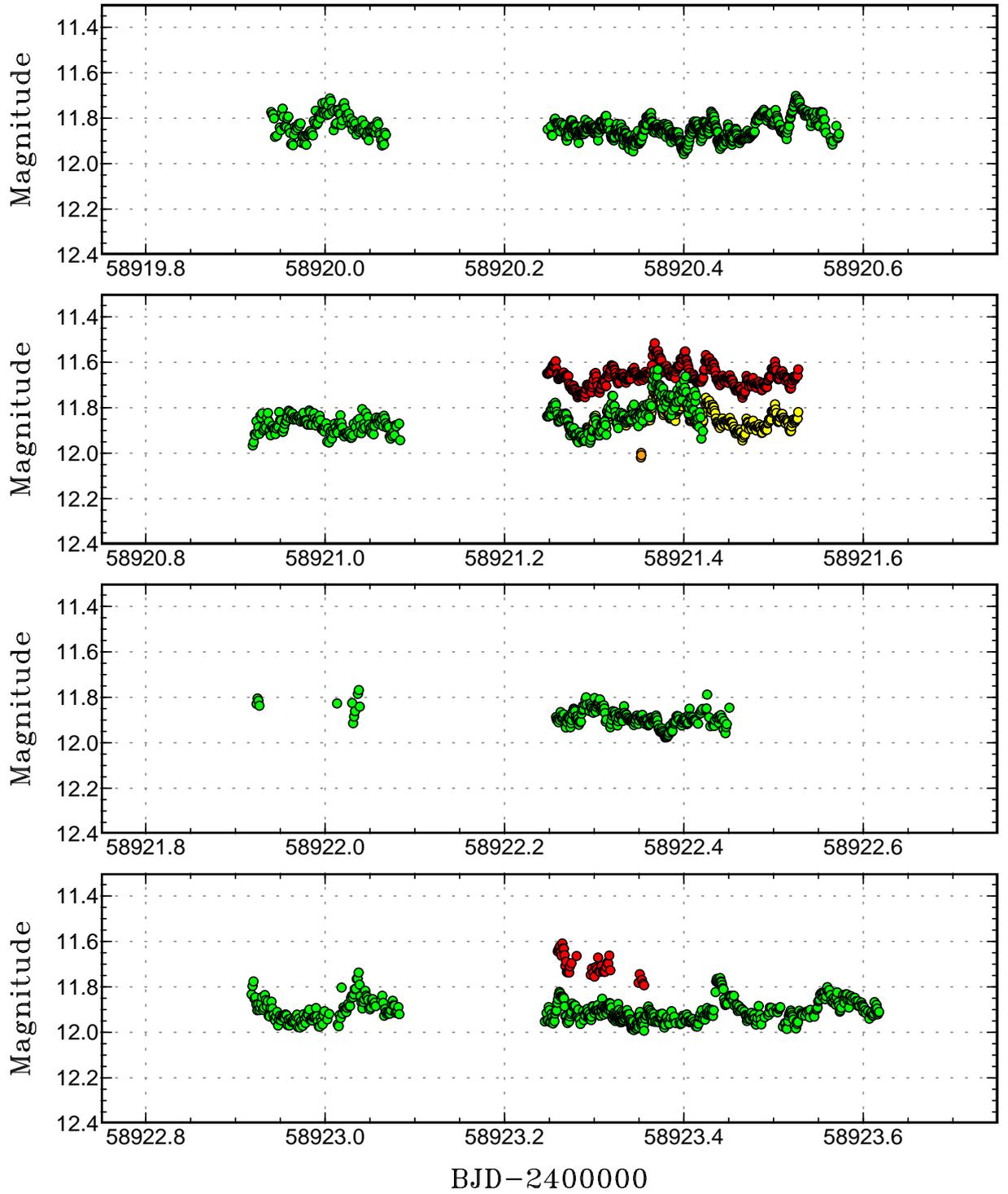}
\caption{
Example of enlarged light curve of ASAS J071404$+$7004.3
using the ground-based observations obtained by our campaign.
The symbols are the same as figure \ref{fig:j0714vsnet}.
}
\label{fig:j0714vsnet2}
\end{center}
\end{figure*}

\begin{figure*}
  \begin{center}
    \includegraphics[width=16cm]{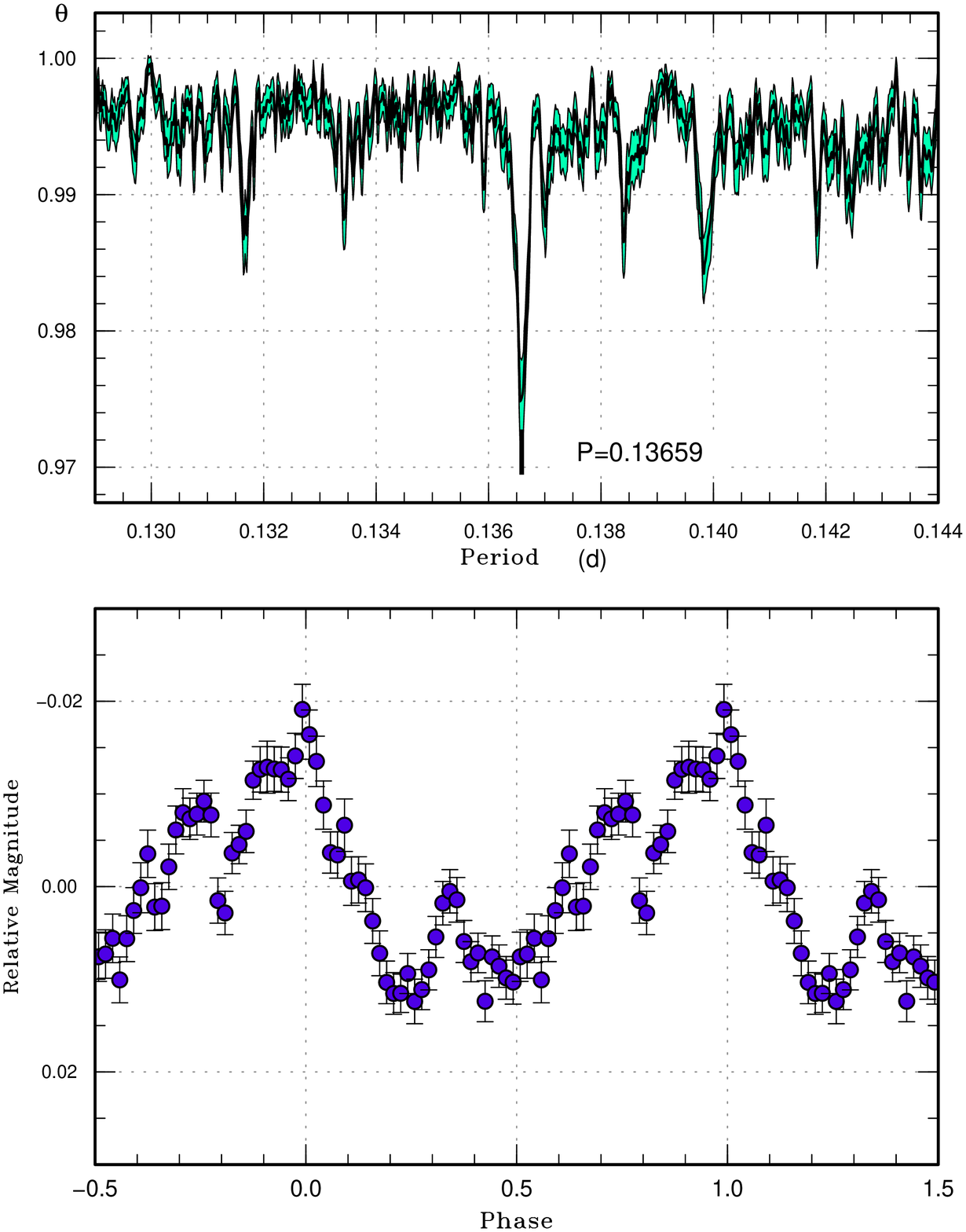}
  \end{center}
  \caption{Period analysis of our campaign in 2020 February--April.
  (Upper): We analyzed 100 samples which randomly contain 50\% of
  observations, and performed the PDM analysis for these samples.
  The bootstrap result is shown as a form of 90\% confidence intervals
  in the resultant PDM $\theta$ statistics.
  (Lower): Orbital variation.
  }
  \label{fig:j0714pdm}
\end{figure*}

We obtained more than 25000 measurements between
2020 February 7 and 2020 April 14.  Although some observations
are apparently common to the ``intensive photometric monitoring of
the system using small-aperture telescopes throughout February and
March 2020'' described in \citet{ini22j0714}, the majority
of the data were obtained by VSOLJ observers,
the Crimean Astrophysical Observatory and the Kyoto University team,
which are not present in the AAVSO International Database\footnote{
  $<$https://www.aavso.org$>$.
} (figures \ref{fig:j0714vsnet} and \ref{fig:j0714vsnet2}).
\citet{ini22j0714} did not show any result of
period analysis of these ground-based observations.
This interval corresponded to the standstill 
(figure \ref{fig:j0714lc3}, the second panel)
and we only searched for periodicities around the orbital period.
We used locally-weighted polynomial regression (LOWESS: \cite{LOWESS})
to remove the long-term trend and applied
the Phase Dispersion Minimization (PDM, \cite{PDM}) method.
The errors of periods by the PDM method were
estimated by the methods of \citet{fer89error} and \citet{Pdot2}.
This analysis (figure \ref{fig:j0714pdm}) detected
an orbital period of 0.136589(5)~d, which agrees with
the value by \citet{ini22j0714} to better than 0.0005\%.

\subsection{Period analysis of TESS data}

\begin{figure*}
  \begin{center}
    \includegraphics[width=16cm]{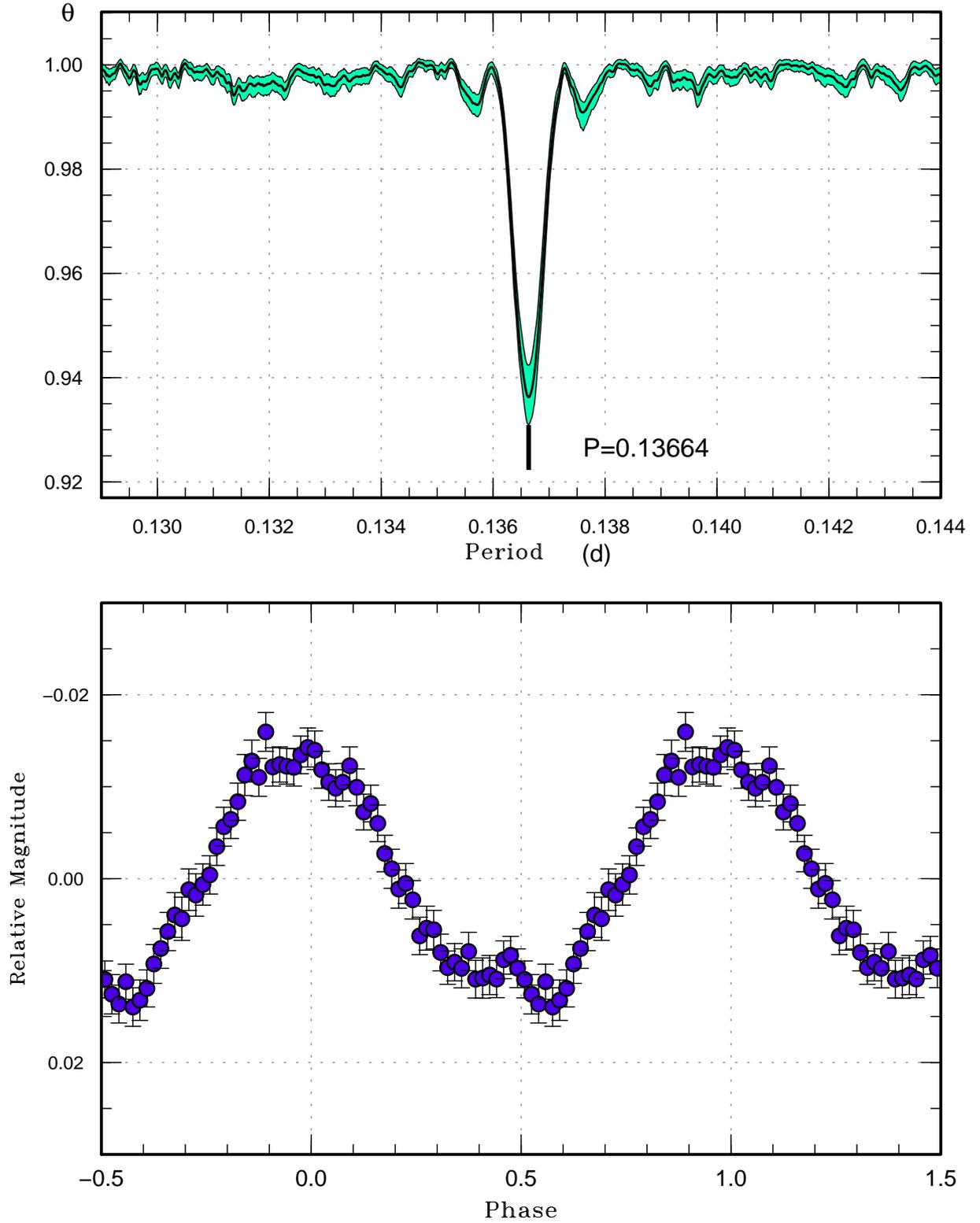}
  \end{center}
  \caption{Period analysis of TESS data between BJD 2458842
  (2019 December 25) and 2458868 (2020 January 20).
  These observations correspond to the long standstill phase.
  (Upper): PDM analysis.
  (Lower): Orbital variation.
  }
  \label{fig:j0714tess1}
\end{figure*}

\begin{figure*}
  \begin{center}
    \includegraphics[width=16cm]{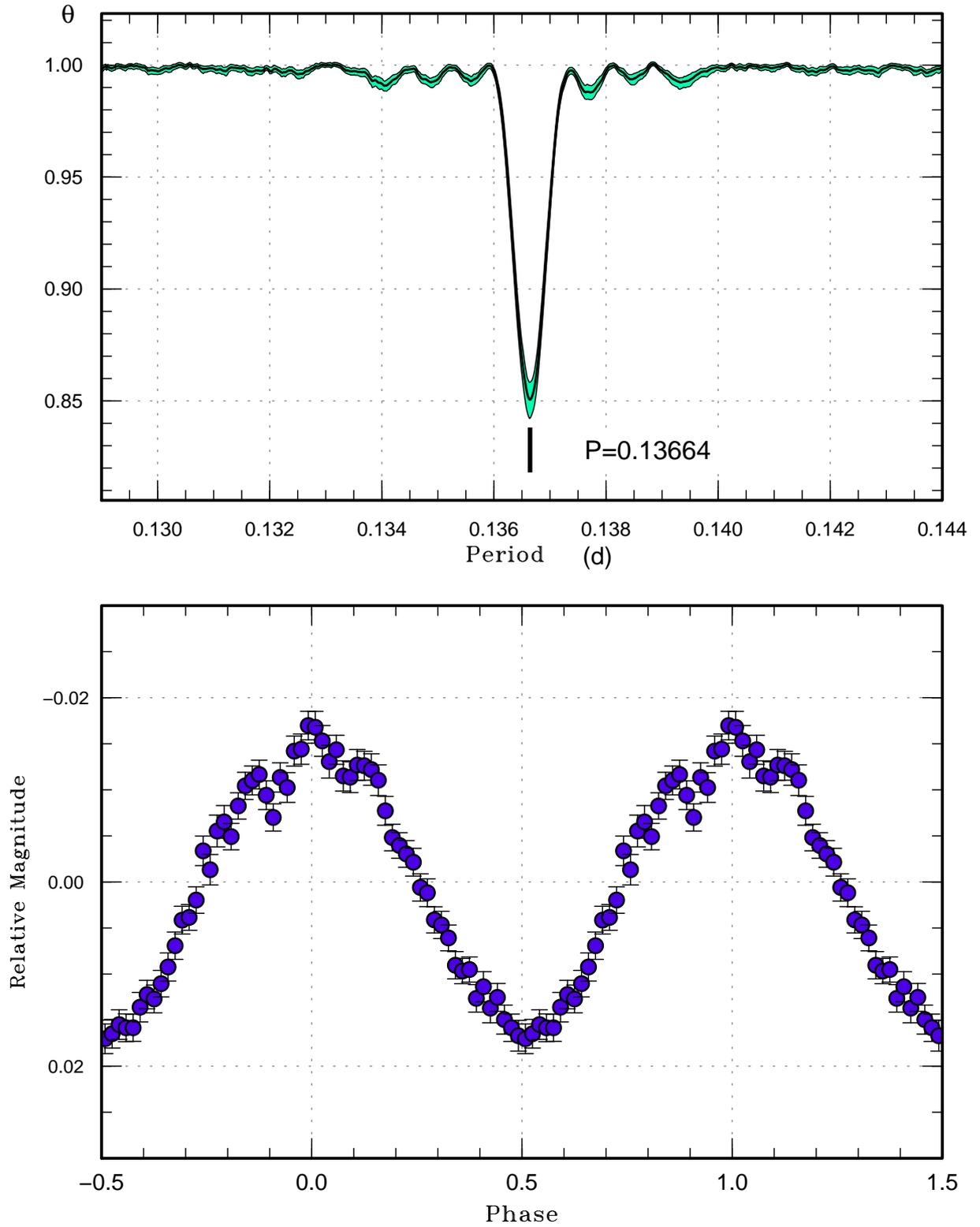}
  \end{center}
  \caption{Period analysis of TESS data between BJD 2459010
  (2020 June 9) and 2459035 (2020 July 4).
  These observations covered the outburst that terminated
  the long standstill phase. 
  (Upper): PDM analysis.
  (Lower): Orbital variation.
  }
  \label{fig:j0714tess2}
\end{figure*}

\begin{figure*}
  \begin{center}
    \includegraphics[width=16cm]{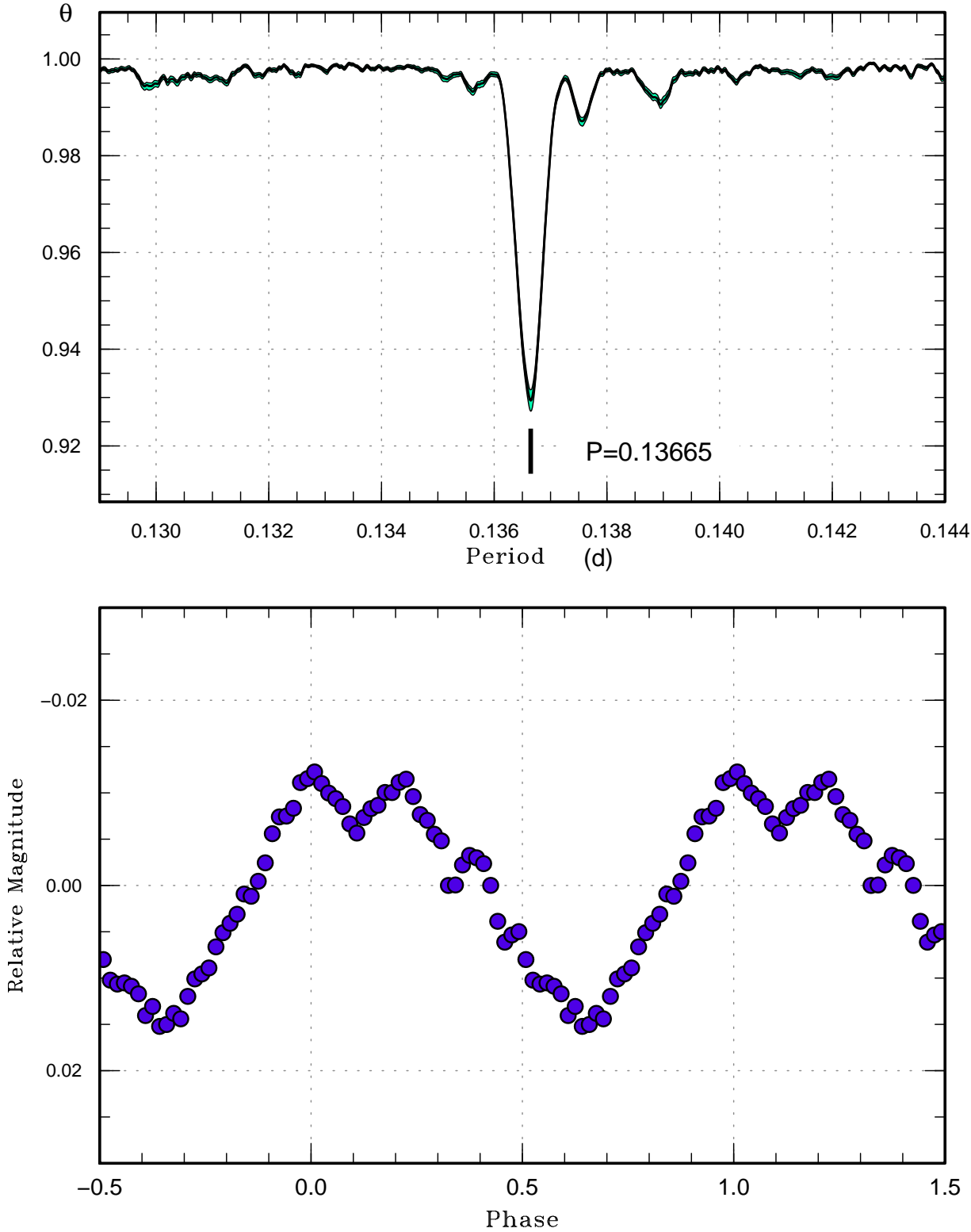}
  \end{center}
  \caption{Period analysis of TESS data between BJD 2459390
  (2021 June 25) and 2459418 (2021 July 23).
  These observations covered one complete outburst and
  one outburst during decline of the heartbeat phase.
  (Upper): PDM analysis.
  (Lower): Orbital variation.
  }
  \label{fig:j0714tess3}
\end{figure*}

\begin{figure*}
  \begin{center}
    \includegraphics[width=16cm]{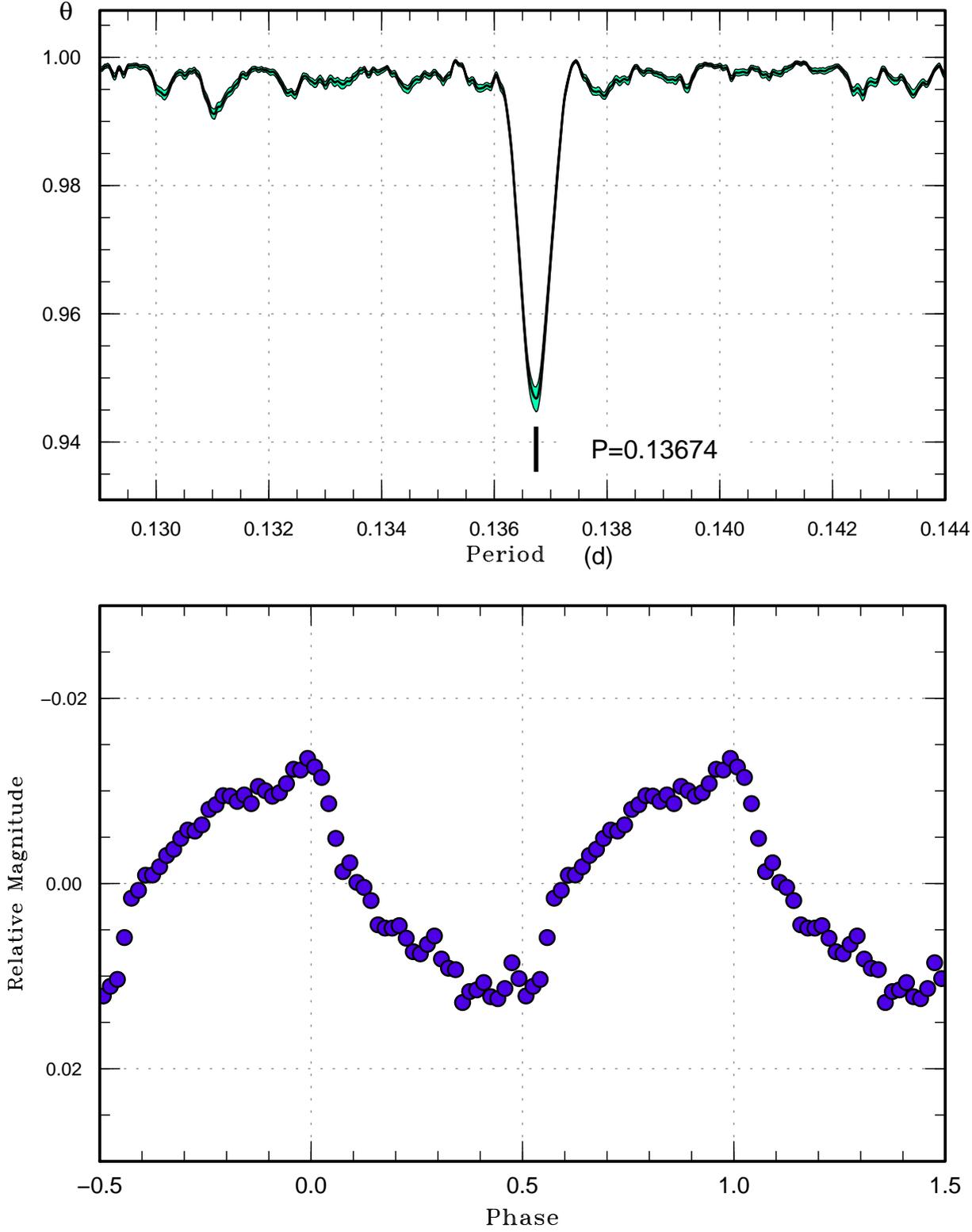}
  \end{center}
  \caption{Period analysis of TESS data between BJD 2459579
  (2021 December 23) and 2459606 (2022 January 27).
  These observations correspond to a slightly variable
  standstill phase.
  (Upper): PDM analysis.
  (Lower): Orbital variation.
  }
  \label{fig:j0714tess4}
\end{figure*}

   TESS observations have four segments.  We performed PDM
analysis as in the same way for the ground-based campaign
(figures \ref{fig:j0714tess1}, \ref{fig:j0714tess2},
\ref{fig:j0714tess3} and \ref{fig:j0714tess4}).
All segments showed only the orbital signal.
We can conclude that that there were neither positive nor
negative superhumps regardless of the outbursting phase.
This result excludes (at least in this object)
a possible suggestion that a tilted disk might be responsible
for the IW And-type phenomenon \citep{kim20kic9406652}.
It has been pointed out that a model involving
a tilted disk could not reproduce the variation of
the disk radius observed in the IW And star KIC 9406652
\citep{kim21kic9406652}.  The absence of negative superhumps
has also been reported in other IW And stars
(IM Eri: \cite{kat20imeri}; ST Cha: \cite{kat21stchaporb};
HO Pup: \cite{lee21hopup}).

\begin{figure*}
  \begin{center}
    \includegraphics[width=15cm]{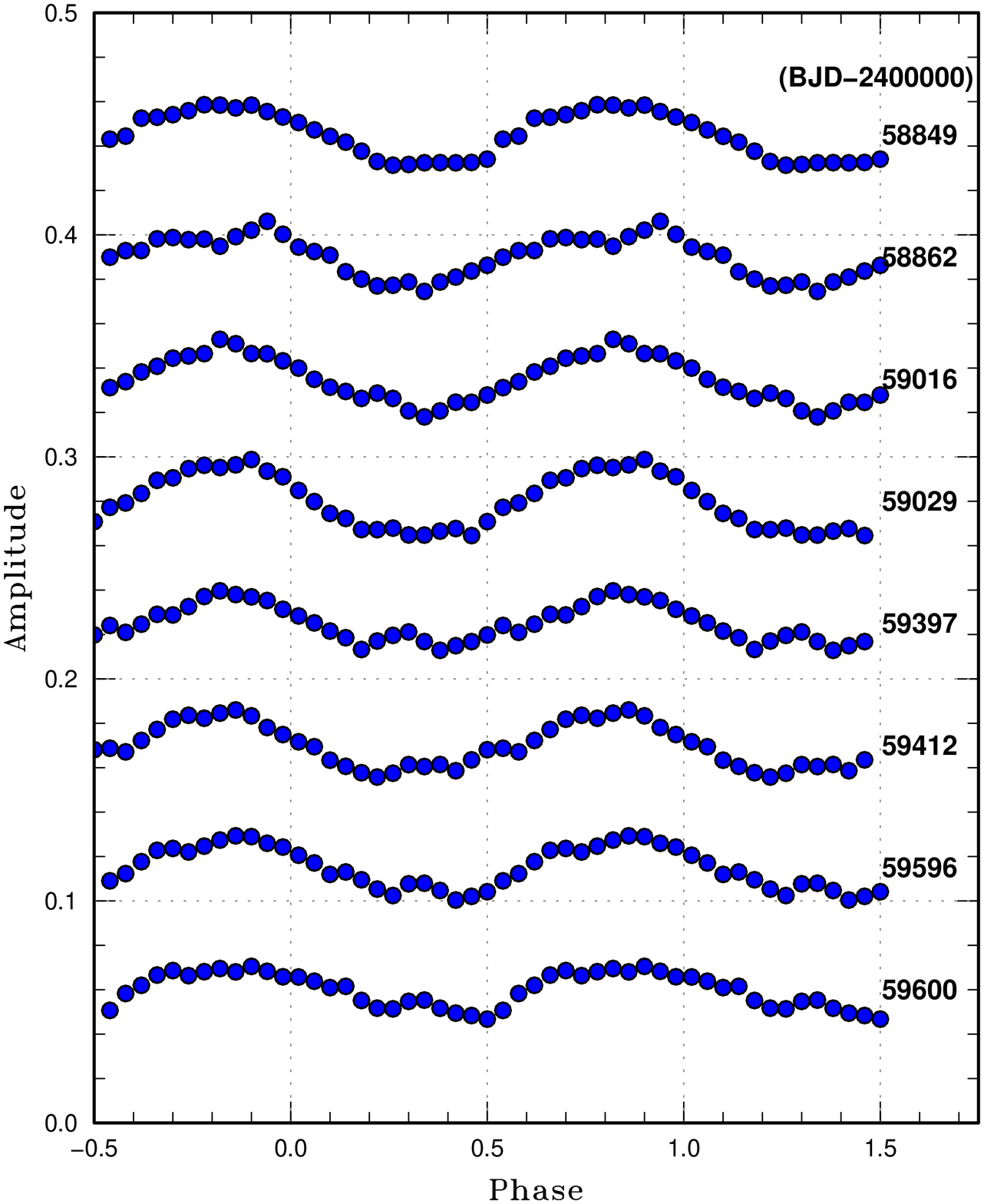}
  \end{center}
  \caption{Variation of the orbital profile in TESS data.
  The data with 14-d segments centered on the dates
  given to the right of the averaged light curves were
  used.  The epoch and period were from \citet{ini22j0714}.
  }
  \label{fig:j0714tessprof}
\end{figure*}

The profile of the orbital variation slightly varied.
During the long standstill and one outburst, the profile
was almost sinusoidal
(figures \ref{fig:j0714tess1}, \ref{fig:j0714tess2}).
During another outburst cycle, the maximum of the orbital
hump had double peaks (figure \ref{fig:j0714tess3}).
During the most recent standstill with some variability,
the orbital profile became asymmetric
(figure \ref{fig:j0714tess4}).  There were no apparent
changes in the PDM $\theta$ diagrams and these changes
were not a result of positive/negative superhumps.
Orbital profiles from the TESS observations are shown
in figure \ref{fig:j0714tessprof}.  It appears that
the phases of maxima and minima varied slightly.
These changes may reflect the changes in the state of
the accretion disk or the accretion geometry, and need
to be studied in detail.

There were no ``embedded precursors'' as observed in
ST Cha \citep{kat21stcha}.  There was no enhancement
of the orbital signal nor appearance of a long period
as observed in a long outburst of the SS Cyg star V363 Lyr
\citep{kat21v363lyr}.

We summarize the values of orbital periods in table
\ref{tab:porb}.  We obtained the photometric orbital
period from the combined TESS data to be 0.1366476(3)~d,
which is in very good agreement with the one obtained
by the radial-velocity study by \citet{ini22j0714}.

\begin{table}
\caption{Orbital period of ASAS J071404$+$7004.3}\label{tab:porb}
\begin{center}
\begin{tabular}{cccc}
\hline
Data & Interval (JD$-$2400000) & Period (d) & Source \\
\hline
Radial velocity & 56597--59197 & 0.1366454(1) & \citet{ini22j0714} \\
TESS (1) & 58842--58868 & 0.136635(10) & this paper \\
TESS (2) & 59010--59035 & 0.136644(7) & this paper \\
TESS (3) & 59390--59418 & 0.136648(3) & this paper \\
TESS (4) & 59579--59606 & 0.136741(4) & this paper \\
TESS combined & 58842--59606 & 0.1366476(3) & this paper \\
VSNET campaign & 58887--58954 & 0.136589(5) & this paper \\
\hline
\end{tabular}
\end{center}
\end{table}

\subsection{Are standstills brighter than dwarf nova-type states?}

\begin{figure*}
  \begin{center}
    \includegraphics[width=15cm]{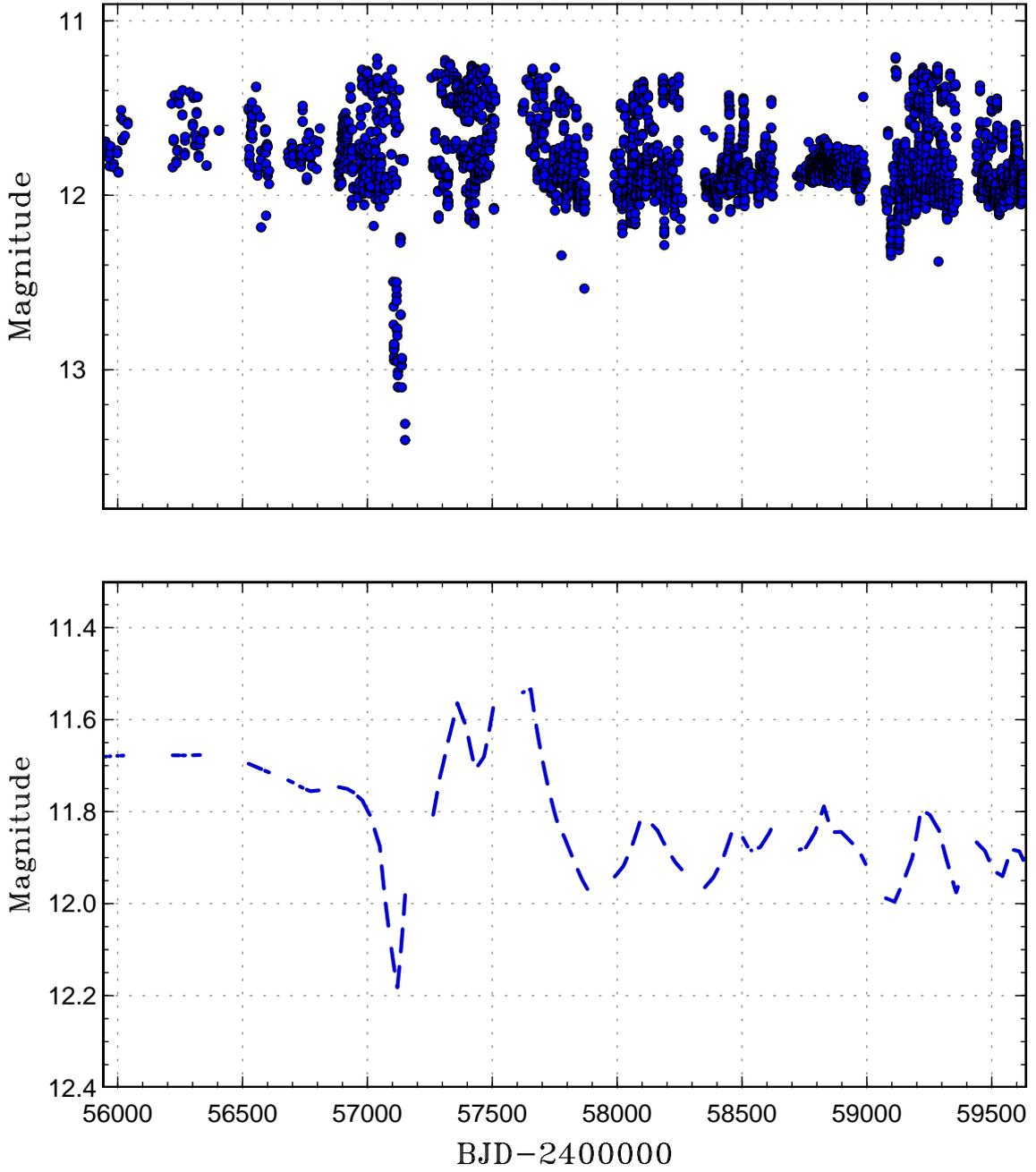}
  \end{center}
  \caption{(Upper): ASAS-SN light curve of ASAS J071404$+$7004.3.
  (Lower): Trend determined by LOWESS.  A smoothing parameter
  of $f$=0.06 was used.}
  \label{fig:j0714ave}
\end{figure*}

   In figure \ref{fig:j0714ave}, we show the trend of mean
magnitudes (averaged in flux) as we did for
IX Vel \citep{kat21ixvel} and ST Cha \citep{kat21stchaporb}.
The result shows that the 2019--2020 standstill was not
brighter than the dwarf nova-type states (IW And-type states)
before and after this.  This indicates that either standstill
or IW And state does not occur as the result of changing
mass-transfer rates.  This conclusion is the same as
in \citet{ini22j0714}.  It would be interesting to note
that dwarf nova-type state with relatively large amplitudes
in 2015--2016 (BJD 2457260--2457700) occurred when
the mean magnitude was the brightest.  This phenomenon
challenges the classical (and widely accepted)
interpretation that standstills in Z Cam stars occur
when the mass-transfer rates are high \citep{mey83zcam}.

\subsection{Notes on interpretations in \citet{ini22j0714}}

\citet{ini22j0714} suggested that there is an outer part
of the disk that occasionally becomes sufficiently cool for
hydrogen to recombine to explain the dwarf nova-like
variations of ASAS J071404$+$7004.3 (this is an explanation
for dwarf-nova outbursts, although \citet{ini22j0714}
did not adopt the dwarf-nova classification for
this object) rather than periodic mass-transfer variations
from the secondary as modeled by \citet{ham14zcam}.
\citet{ini22j0714} stated that this outburst starts
at the inner edge of this area and subsequently spreads
outwards.  This picture is essentially what was proposed for
IW And stars in \citet{kat19iwandtype} (it was written as
``the standstill in these objects is somehow maintained in
the inner part of the disk and the thermal instability
starting from the outer part of the disk terminates
the standstill to complete the cycle'' in the abstract).
It is, however, difficult to consider a disk whose outer
part becomes cool for a long time when the mass-transfer
hits the outer edge of the disk.
\citet{kim20iwandmodel} considered
a tilted disk in which the majority of mass stream hits
the inner disk and the outer disk is cool most of the time.
Although model simulations by \citet{kim20iwandmodel}
reproduced a limit cycle similar to IW And stars,
they could not reproduce the observed disk radius variation
\citep{kim20kic9406652} and the frequent occurrence of heating
waves in the cool part of the disk was somewhat different
from what are observed in IW And stars.
The existence of the cool region in the disk of IW And
stars during (quasi-)standstills has been confirmed
by \Shibataprep\, and the picture
by \citet{kat19iwandtype} and \citet{ini22j0714} appear
to be correct.  We still need an explanation why
the outer disk can remain cool in the presence of
mass-transfer reaching the outer part of the disk.

\citet{ini22j0714} also concluded that the emission lines
in ASAS J071404$+$7004.3 could be reproduced by considering
the wind.  It would be worth noting that \citet{tam22v455and}
considered the same kind of wind during the superoutburst of
the WZ Sge star V455 And and reproduced singly peaked
emission lines as observed, which are not expected for this
high-inclination CV.

\section*{Acknowledgements}

This work was supported by JSPS KAKENHI Grant Number 21K03616.
The author is grateful to the ASAS-SN team for making their data
available to the public.
This research has made use of the AAVSO Variable Star Index
and NASA's Astrophysics Data System.

\section*{List of objects in this paper}
\xxinput{objlist.inc}

\section*{References}

We provide two forms of the references section (for ADS
and as published) so that the references can be easily
incorporated into ADS.

\renewcommand\refname{\textbf{References (for ADS)}}

\newcommand{\noop}[1]{}\newcommand{\hyphalt}{-}

\xxinput{j0714aph.bbl}

\renewcommand\refname{\textbf{References (as published)}}

\xxinput{j0714.bbl.vsolj}

\end{document}